\documentclass{elsart}
\newcommand{\BRS}{BR scaling}

\begin{document}
\runauthor{Matsuzaki and Tanigawa}
\begin{frontmatter}
\title{Effects of Meson Mass Decrease \\
on Superfluidity in Nuclear Matter}
\author[fue]{\underline{Masayuki Matsuzaki}\thanksref{emm}}
\author[kyushu]{Tomonori Tanigawa\thanksref{tmm}}

\address[fue]{Department of Physics, Fukuoka University of Education, 
Munakata, \\
Fukuoka 811-4192, Japan}
\address[kyushu]{Department of Physics, Kyushu University, 
Fukuoka 812-8581, Japan}
\thanks[emm]{Electronic address: matsuza@fukuoka-edu.ac.jp}
\thanks[tmm]{Electronic address: tomo2scp@mbox.nc.kyushu-u.ac.jp}
\begin{abstract}
We calculate the $^1S_0$ pairing gap in nuclear matter 
by adopting the ``in-medium Bonn potential" proposed by Rapp 
et al. [e-print nucl-th/9706006], which 
takes into account the in-medium meson mass decrease, as the 
particle-particle interaction in the gap equation. The resulting gap is 
significantly reduced in comparison with the one obtained by adopting the 
original Bonn potential.
\end{abstract}
\end{frontmatter}
\newpage

 Superfluidity in infinite hadronic matter has long been studied mainly in 
neutron matter from a viewpoint of neutron-star physics such as its cooling 
rates. As a way of description, relativistic models 
are attracting attention in addition to traditional non-relativistic 
nuclear many-body theories. Since Chin and Walecka 
succeeded in reproducing the saturation property of symmetric nuclear matter 
within the mean-field theory (MFT)~\cite{cw1}, quantum 
hadrodynamics (QHD) which is an effective field theory of hadronic degrees of 
freedom has described not only infinite matter but also finite spherical, 
deformed and rotating nuclei successfully with various 
approximations~\cite{rev1,rev2}.  These successes indicate that the 
particle-hole (p-h) channel in QHD is realistic. In contrast, relativistic 
nuclear structure calculations with pairing done so far have been using 
particle-particle (p-p) interactions borrowed from non-relativistic models 
such as Gogny force. 
Aside from practical successes of this kind of calculations, the p-p channel 
in QHD itself is a big subject which has just been started to study.

 The first study of this direction was done by Kucharek and Ring~\cite{kr}. 
They adopted, as the particle-particle interaction ($v_\mathrm{pp}$) in the 
gap equation, a one-boson-exchange interaction with the ordinary 
relativistic MFT parameters, which gave the saturation 
under the no-sea approximation.  The resulting maximum gap 
was about three times larger than the accepted values in the non-relativistic 
calculations.~\cite{nrel1,baldo,nrel2,nrel3,nrel4}. 
Various modifications to improve this result were 
proposed~\cite{rel1,rel2,mm} but this has still been an open problem. 
One of such modifications is to include the p-h and the N-$\bar\mathrm{N}$ 
polarizations, the former of which is effective 
for reducing the gap in the non-relativistic models~\cite{pol1,pol2}.

 From a different viewpoint, Rummel and Ring~\cite{rr,rev1} adopted the Bonn 
potential~\cite{bonn}, which was constructed so as to reproduce the phase 
shifts of nucleon-nucleon scattering in free space, as $v_\mathrm{pp}$ and 
obtained pairing gaps consistent with the non-relativistic studies. 
Since the single particle states are determined by the MFT also in this 
calculation, explicit consistency between the p-h and the p-p channels is 
abandoned. However, assuming that the MFT simulates the 
Dirac-Brueckner-Hartree-Fock (DBHF) calculation using the Bonn potential, 
we can consider that the consistency still holds implicitly. 
A possible extension of this study is to take into account the in-medium 
changes of the properties of the mesons which mediate the inter-nucleon 
forces. Among them, the change of the mass, that is, the momentum-independent 
part of the self-energy, has been discussed in terms of the Brown--Rho (BR) 
scaling~\cite{br}, 
the QCD sum rules~\cite{hatsu}, and some other models both in the quark level 
and in the hadron level~\cite{jean,qmc}. Although there still has been 
theoretical controversy~\cite{nsd}, some experiments look to support the 
vector meson 
mass decrease~\cite{exp}. In addition, the momentum-dependent part also 
attracts attention recently~\cite{ele,fri,lee,peters}. In the present work, 
we assume only the momentum-independent vector meson mass decrease.
A simple way to take into account this meson mass decrease in the meson 
exchange interactions is to assume the \BRS.
Actually Rapp et al.~\cite{rapp} showed, based on this, 
that the saturation property of symmetric nuclear matter and the mass decrease 
of the vector mesons were compatible. So we adopt their ``in-medium Bonn 
potential" as $v_\mathrm{pp}$ in the gap equation. Since the gap equation 
takes the form such that the short range correlation is 
involved~\cite{cor1,cor2,nrel2}, we use this interaction in the gap equation 
without the iteration to construct the $G$-matrix.

 As described in ref.~\cite{kr}, meson fields also have to be treated 
dynamically beyond the MFT to incorporate the pairing field via the anomalous 
(Gor'kov) Green's functions~\cite{gor}. The resulting 
Dirac-Hartree-Fock-Bogoliubov 
equation reduces to the ordinary BCS equation in the infinite matter case. 
We start from a model Lagrangian for the nucleon, the $\sigma$ 
boson, and the $\omega$ meson,
\begin{eqnarray}
\mathcal{L}&=&\bar\psi(i\gamma_\mu\partial^\mu-M)\psi
\nonumber\\
 &+&{1\over2}(\partial_\mu\sigma)(\partial^\mu\sigma)
  -{1\over2}m_\sigma^2\sigma^2
  -{1\over4}F_{\mu\nu}F^{\mu\nu}+{1\over2}m_\omega^2\omega_\mu\omega^\mu
%\nonumber\\
% &-&{1\over4}\vec{R}_{\mu\nu}\cdot\vec{R}^{\mu\nu}
%  +{1\over2}m_\rho^2\vec{\rho}_\mu\cdot\vec{\rho}^\mu
\nonumber\\
 &+&g_\sigma\bar\psi\sigma\psi-g_\omega\bar\psi\gamma_\mu\omega^\mu\psi ,
%  -g_\rho\bar\psi\gamma_\mu\vec{\tau}\cdot\vec{\rho}^\mu\psi ,
%\nonumber\\
% &-&{1\over{3!}}\kappa\sigma^3-{1\over{4!}}\lambda\sigma^4 ,
\nonumber\\
F_{\mu\nu}&=&\partial_\mu\omega_\nu-\partial_\nu\omega_\mu .
%\nonumber\\
%\vec{R}_{\mu\nu}&=&\partial_\mu\vec{\rho}_\nu
%                  -\partial_\nu\vec{\rho}_\mu ,
\end{eqnarray}
The actual task is to solve the coupled 
equations for the effective nucleon mass and the $^1S_0$ pairing gap:
\begin{eqnarray}
  M^\ast&=&M-{{g_\sigma^2}\over{m_\sigma^2}}{\gamma\over{2\pi^2}}
  \int_0^{\Lambda_\mathrm{c}} 
  {M^\ast\over\sqrt{{\bf k}^2+M^{\ast\,2}}}v^2(k)k^2dk ,
%  \nonumber\\
%  &+&{{\kappa}\over{2g_\sigma m_\sigma^2}}(M-M^\ast)^2
%   +{{\lambda}\over{6g_\sigma^2 m_\sigma^2}}(M-M^\ast)^3 ,
  \nonumber\\
  \Delta(p)&=&-{1\over{8\pi^2}}
  \int_0^{\Lambda_\mathrm{c}} \bar v_\mathrm{pp}(p,k)
         {\Delta(k)\over\sqrt{(e_k-e_{k_\mathrm{F}})^2+\Delta^2(k)}}k^2dk ,
  \nonumber\\
  v^2(k)&=&{1\over2}\left(1-
   {{e_k-e_{k_\mathrm{F}}}\over\sqrt{(e_k-e_{k_\mathrm{F}})^2+\Delta^2(k)}}
                    \right) ,
  \nonumber\\
  e_k&=&\sqrt{{\bf k}^2+M^{\ast\,2}}+g_\omega\langle\omega^0\rangle ,
\end{eqnarray}
where $\bar v_\mathrm{pp}(p,k)$ is an anti-symmetrized matrix element of the 
adopted p-p interaction,
\begin{eqnarray}
& &\bar v_\mathrm{pp}({\bf p},{\bf k})
\nonumber\\
& &=\langle{\bf p}s',\tilde{{\bf p}s'}\vert v_\mathrm{pp}\vert
           {\bf k}s,\tilde{{\bf k}s}\rangle
   -\langle{\bf p}s',\tilde{{\bf p}s'}\vert v_\mathrm{pp}\vert
           \tilde{{\bf k}s},{\bf k}s\rangle ,
\end{eqnarray}
with an instantaneous approximation (energy transfer $=$ 0) and an integration 
with respect to the angle between ${\bf p}$ and ${\bf k}$ to project out the 
$S$-wave component, and $\langle\omega^0\rangle$ is determined by the baryon 
density. Here we consider only symmetric nuclear matter ($\gamma=$ 4) and 
neutron matter ($\gamma=$ 2). Conceptually the numerical integrations run to 
infinity in the present model where the finiteness of the hadron size is 
considered in $v_\mathrm{pp}$. Numerically, however, we introduce a cut-off 
$\Lambda_\mathrm{c}$; its value is chosen to be 20 fm$^{-1}$ so that the 
integrations converge. If we adopt the Bonn potential as $v_\mathrm{pp}$ and 
include the $\rho$ meson and the cubic and 
quartic self-interactions of $\sigma$ with choosing the NL1 parameter 
set~\cite{rein} for the mean field, these equations reproduce 
the results of ref.~\cite{rr}. Here we note that there are pros and cons 
as for the self-interaction terms~\cite{bb,furn}.
First we developed a computer code to calculate the Bonn potential for this 
channel from scratch, and later confirmed that our code 
reproduced outputs of the one in ref.~\cite{mach}.

 Then in the present work we adopt the in-medium Bonn potential of Rapp 
et al.~\cite{rapp} ~although this is, as yet, unpublished. 
To construct this potential, they applied the 
\BRS ~after replacing the $\sigma$ boson in the original Bonn-B 
potential with the correlated and the uncorrelated $2\pi$ exchange processes. 
Then they parameterized the obtained nucleon-nucleon interaction by three 
scalar bosons $\sigma_1$, $\sigma_2$ and ``rest-$\sigma$", in addition to 
$\pi$, $\eta$, $\rho$, $\omega$ and $\delta$. Here $\sigma_1$ and  $\sigma_2$ 
with density-dependent masses and coupling constants simulate the correlated 
$2\pi$ processes and the ``rest-$\sigma$" simulates the uncorrelated ones. 
Therefore the actual tasks are adding two extra scalar bosons to the original 
Bonn-B potential and applying the \BRS,
\begin{equation}
 {{M^\ast}\over{M}}
={{m_{\rho,\omega}^\ast}\over{m_{\rho,\omega}}}
={{\Lambda_{\rho,\omega}^\ast}\over{\Lambda_{\rho,\omega}}}
=1-C{\rho\over{\rho_0}} ,
\end{equation}
where $\rho_0$ is the normal nuclear matter density, $\Lambda_{\rho,\omega}$ 
are the cut-off masses in the form factors of the meson-nucleon vertices, 
which is related to the hadron size. The scaling parameter $C$ is chosen to be 
0.15.  Aside from a slight tuning of the coupling constant of $\delta$, other 
parameters are the same as those in the original Bonn-B potential. All the 
potential parameters are presented in Tables I and II in ref.\cite{rapp}. 
Note that this scaling applies only to the quantities in $v_\mathrm{pp}$, 
since the MFT with its effective nucleon mass guarantees the saturation and 
we confirmed that the effects of pairing on the bulk properties are negligible 
except at very low densities in the present model. 
As for the $\pi$-N coupling in the potential, we adopt the pseudovector one 
conforming to the suggestion of chiral symmetry~\cite{wein} and actually 
the pseudoscalar one in the Bonn potential was shown to lead to 
unrealistically attractive contributions in the DBHF calculation~\cite{pv}.

 The result is presented in Fig.1. This shows that applying the \BRS ~reduces 
the gap significantly in comparison with the original Bonn-B potential.
The potentials themselves are given in Fig.2. The relation between the 
reduction of the gap and the change in the potential, i.e., the shift to 
lower momenta, can be understood as follows:
As discussed in refs.~\cite{baldo,rr}, $\Delta(k)$ determined by the second 
equation of (2) exhibits a nodal structure similar to 
$v_\mathrm{pp}(k_\mathrm{F},k)$ 
as functions of $k$ with the opposite sign and some possible deviation of 
zeros. Consequently both the low-momentum attractive part where $\Delta(k)>0$ 
and the high-momentum repulsive part where $\Delta(k)<0$ give positive 
contributions to $\Delta(k_\mathrm{F})$, and Fig.2 shows that the both of 
these contributions are reduced in the case of the in-medium Bonn potential. 
This is the main reason why $\Delta(k_\mathrm{F})$ is reduced by applying the 
\BRS. Actually we confirmed that this is mainly due to the mass decrease of 
the vector mesons, not of the nucleon, as shown in Fig.3 although also the 
latter produces some reduction of the gap. Note here that the 
saturation in the DBHF calculation is guaranteed only 
when both $M$ and $m_{\rho,\omega}$ are scaled. The in-medium meson mass 
decrease is brought about by the N-$\bar\mathrm{N}$ polarization in hadronic 
models. Although its explicit inclusion in the gap equation has not been 
reported yet, it is conjectured in ref.~\cite{mm} that 
its inclusion simultaneously with the p-h one will reduce the gap as that of 
the p-h one in the  non-relativistic models~\cite{pol1,pol2}. In this sense, 
the present result that applying the \BRS ~reduces the gap is consistent with 
this conjecture, although the reduction in refs.\cite{pol1,pol2}~ is more 
drastic than that in the present work. 
An additional element is that the cut-off masses in the form factors of the 
meson-nucleon vertices are also scaled. Since they appear in the form
\begin{equation}
{{\Lambda_\alpha^2-m_\alpha^2}\over{\Lambda_\alpha^2+{\bf q}^2}} ,
\end{equation}
where ${\bf q}$ is the momentum transfer~\cite{bonn}, simultaneous 
reductions of $\Lambda_\alpha$ and $m_\alpha$ ($\alpha=\rho, \omega$), lead 
to a further reduction of the repulsion due to the vector mesons at 
large-$\vert{\bf q}\vert$. Since the higher the background density becomes 
the more nucleons feel the high-momentum part of $v_\mathrm{pp}$, 
the in-medium reduction of $\Lambda_{\rho,\omega}$ leads to an additional 
reduction of the gap in the high-density region.

 Finally, use of the NL1 parameter set for the 
mean field brings negligible changes and the result for neutron matter is very 
similar to those presented here for symmetric nuclear matter.

 To summarize, we adopted the in-medium Bonn potential, proposed by Rapp 
et al., which takes into account the in-medium meson mass decrease in a 
simple form and is compatible with the nuclear matter saturation in the DBHF 
calculation, as the particle-particle interaction in the gap equation. 
The resulting pairing gap in nuclear matter 
is significantly reduced in comparison with the one obtained by adopting the 
original Bonn potential. Here we should note that the uncertainty in the 
gap value deduced from various studies is still larger than the strength of 
the reduction found in the present work, and therefore, further investigations 
are surely necessary both relativistically and non-relativistically.

\newpage

%\begin{center}
{\bf Figure Captions}
%\end{center}
%\begin{figure}[t]
%\begin{center}
%\epsfig{figure=matsu1.eps,width=8cm}
%\end{center}
%\caption{

Fig.1.~
Pairing gap in symmetric nuclear matter at the Fermi surface as 
functions of the Fermi momentum. Dotted and solid lines indicate the 
results obtained by adopting the Bonn-B and the in-medium Bonn potentials, 
respectively. Note that accuracy is somewhat less around 
$\Delta(k_\mathrm{F})\simeq0$ in our code.
%}
%\label{fig:figa}
%\end{figure}

%\begin{figure}[h]
%\begin{center}
%\epsfig{figure=matsu2.eps,width=10cm}
%\end{center}
%\caption{
Fig.2.~
Matrix element $\bar v_\mathrm{pp}(k_\mathrm{F},k)$ as functions of the 
momentum $k$, with a Fermi momentum $k_\mathrm{F}=$ 0.9 fm$^{-1}$. Dotted and 
solid lines indicate the results obtained by adopting the Bonn-B and the 
in-medium Bonn potentials, respectively.
%}
%\label{fig:figb}
%\end{figure}

%\begin{figure}[h]
%\begin{center}
%\epsfig{figure=matsu3.eps,width=8cm}
%\end{center}
%\caption{
Fig.3.~
Pairing gap in symmetric nuclear matter at the Fermi surface as 
functions of the Fermi momentum. Solid, dotted and dashed lines indicate that 
the results obtained by reducing according to Eq. (4) only the nucleon mass, 
only the vector meson masses, and the both, respectively.
%}
%\label{fig:figc}
%\end{figure}

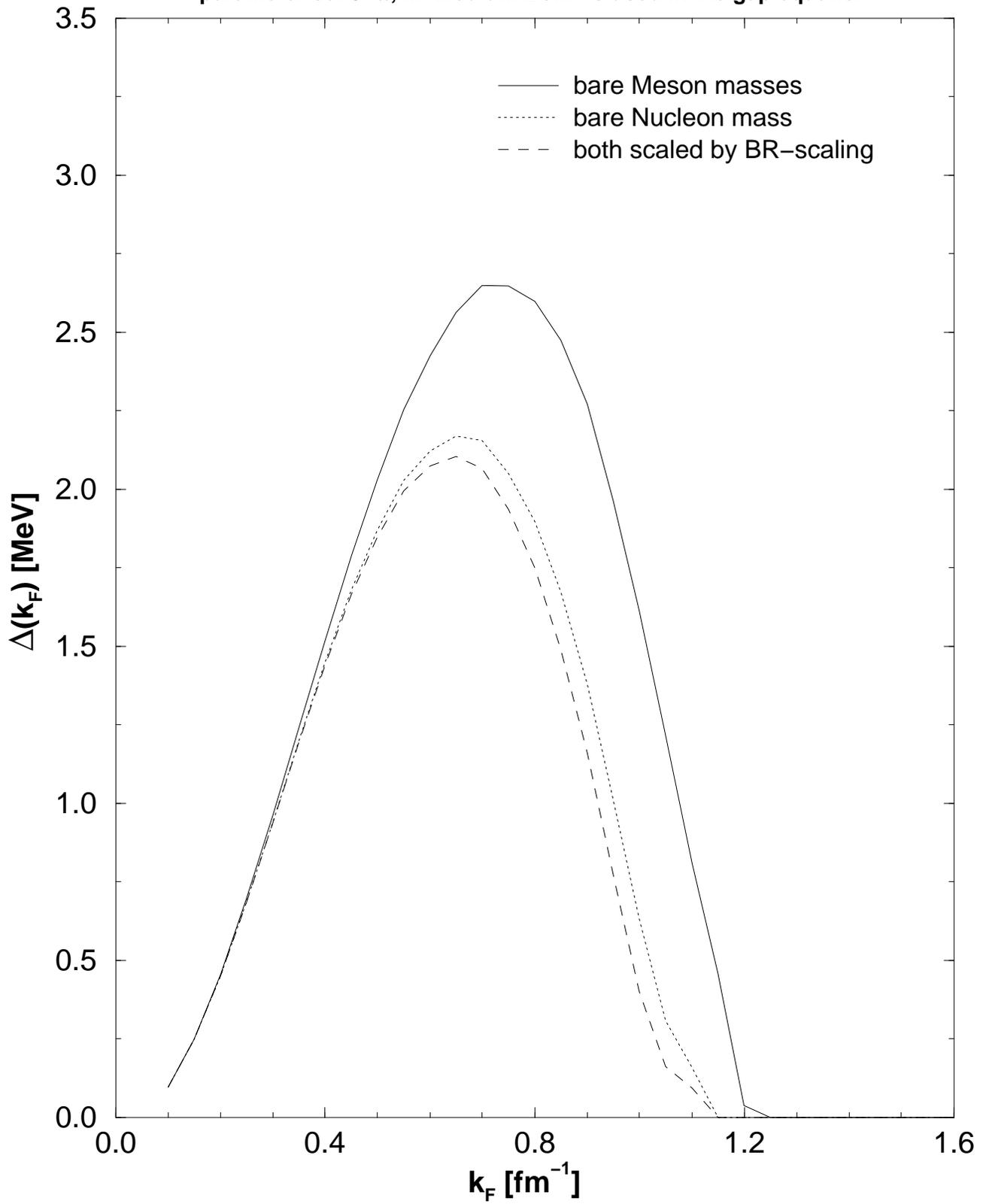
\end{document}